\documentstyle[twocolumn,aps,prd,epsf,epsfig]{revtex}

\begin{document}

\twocolumn[\hsize\textwidth\columnwidth\hsize
\csname@twocolumnfalse%
\endcsname
\draft
\title{Bose--Einstein--Young condensates}
\author{I.E.Mazets}
\address{{\setlength{\baselineskip}{18pt}
Ioffe Physico-Technical
Institute, 194021 St.Petersburg, Russia }} 
\maketitle

\begin{abstract}
We demonstrate a possibility to create a new state of 
ultracold atoms which we call a Bose--Einstein--Young condensate. 
Atoms are supposed to be in different hyperfine state of the same 
isotope. 
The wave function of such a state, although totally symmetric with 
respect to {\it simultaneous} permutation of co-ordinates and spins 
of any pair of atoms, has more complicated structure than a simple 
product of totally symmetric co-ordinate and spin parts. Its 
properties with respect to permutations of {\it only} co-ordinates 
or {\it only} spins are characterized by a particular Young 
diagram, a symbol denoting an irreducible representation of the 
permutation group. \\  
\end{abstract}
\vskip1pc]

Creation of binary mixtures of Bose--Einstein condensates (BECs) of 
ensembles of atoms of the same isotope in different hyperfine states 
\cite{cornell1,cornell2} and spinor condensates in optical traps 
\cite{ketterle1,chapman} is one of the remarkable advances in 
experiments with BECs. The results of these experiments are in 
good agreement with the existing theories of a multicomponent BEC
\cite{hs,hy}. Even if one deals with a spinor BEC, where an 
external magnetic field gives rise to magnetization and, hence, 
non-trivial spin properties, including condensate fragmentation 
\cite{hy}, the translational motion of atoms at 
sufficiently low temperature 
(disregarding small quantum depletion) is essentially the ground state 
of a trap that can be determined by energy minimization. 

However, a question arises, is such a ground state always a final 
state reached at the end of a process of cooling of atoms. In the 
present Rapid Communication, 
we demonstrate that the answer is negative. It is 
possible to find various states those should appear as final states 
of a sympathetic cooling process and are not coupled by 
interatomic interactions to the ground state determined in 
Refs.\cite{hs,hy}.

Due to the Bose-Einstein statistics, the wave function of the atoms 
must be invariant with respect to permutation of both the 
co-ordinates $({\bf r}_i)$ and spins $(\sigma _i)$ 
of any pair of atoms. In general case, it can 
be constructed in different ways, namely, 
\begin{eqnarray}
\Psi ({\bf r}_1,\,...\,,\,{\bf r}_n;\,\sigma _1,\,...\,,\,\sigma _n)
&=&\frac 1{\sqrt{s^{\{ \lambda \} } }} 
\sum _{k=1}^{s^{\{ \lambda \} }}\phi _k^{\{ \lambda \} }
({\bf r}_1,\,...\,,\,{\bf r}_n)  \nonumber  \\ & &
\times \chi _k^{\{  \lambda \}  }(\sigma _1,\,...\,,\,\sigma _n). 
\label{totsym}
\end{eqnarray}
Here $\{ \lambda \}$ denotes an irreducible representation of the 
symmetric group of the order $n$ (i.e., the group of permutations of 
$n$ objects), $n$ is the number of identical atoms in the system, 
and $s^{\{ \lambda \} }$ is the dimension of the 
representation. The theory of symmetric group representations and its 
applications in physics are described, e.g., in classical textbooks 
by Wigner \cite{wigner} and by Lyubarskii \cite{gl}. It is worth to note 
that theory of symmetric group representation is widely used in theory 
of atomic spectra \cite{wybourne} as well as of rotational and 
vibrational spectra of symmetric  polyatomic molecules containing 
identical nuclei. Here we remind briefly the necessary subjects from 
the theory of symmetric group representations. 

Any irreducible representation of the symmetric group of order $n$ 
is characterized by a partition $\{ \lambda \} =\{ \lambda _1,\, 
\lambda _2,\, ...\,,\,\lambda _r\}$, i.e., representation of the 
number $n$ as a sum of integer positive summands arranged in the 
non-increasing order: 
$$
n=\lambda _1+\lambda _2+...+\lambda _r, 
$$
$$
\lambda _1\geq \lambda _2\geq ...\geq \lambda _r>0.
$$
Graphically the partition $\{ \lambda \}$ can be represented by 
a so-called Young diagram consisting of $n$ square boxes arranged 
in rows, the number of boxes in {\it i}-th row is equal to 
$\lambda _i$. To get a basis for this representation, one has to 
associate each co-ordinate or spin variable to one of the boxes. 
Then it is necessary to apply   
 to a co-ordinate or, 
respectively, spin wave function a certain 
symmetrization-antisymmetrization procedure that involves 
particular permutations of the function arguments, 
according to their distribution between the boxes, and 
summation of the obtained terms with the sign of a term 
depending on a given permutation operator. This 
procedure is explicitly described in various texbooks, e.g.,  in 
Ref.\cite{gl}. The number of linearly independent functions obtained 
in such a way starting from different distribution of variables 
between the boxes of the Young diagram 
is the dimension of the representation. It can 
be calculated using the following formula \cite{wybourne}: 
\begin{equation}
s^{\{ \lambda \} }=n!\frac {\prod _{i<i^\prime }(l_i-l_{i^\prime })}
{\prod _il_i!}, 
\label{sl}
\end{equation} 
where $l_i=\lambda _i+r-i$, $i=1,\,2,\,...\,,\,r$. The sum of squares 
of dimensions of all the irreducible representations equals to the 
number of operations in the group: 
\begin{equation}
\sum _{ \{ \lambda \} }s^{\{  \lambda \} \,2}=n!. 
\label{potom}
\end{equation}

A basis of a given representation can be orthogonalized and normalized 
(by means of linear transformation) with respect to the usual quantum 
mechanical definition of the scalar product, implying integration 
over coordinates and summation over spin variables. Thus we get the 
two sets of functions $\phi ^{\{ \lambda \} }_k$ and 
$\chi ^{\{ \lambda \}\,* }_k$, depending on atomic co-ordinates and 
spins, respectively. If we denote the operators of 
permutations of the co-ordinates and spins of the {\it i}-th and 
{\it i}$^\prime $-th atoms by 
$\hat{P}(i,\,i^\prime )$ and 
$\hat{Q}(i,\,i^\prime )$, respectively, the transformation law 
can be written as follows:
\begin{equation}
\hat{P}(i,\,i^\prime )\phi ^{\{ \lambda \} }_k=
\sum _m^{  s^{\{ \lambda \} }  }T_{km}(i,\,i^\prime )
\phi ^{\{ \lambda \} }_m, 
\label{tp}
\end{equation}
\begin{equation}
\hat{Q}(i,\,i^\prime )\chi ^{\{ \lambda \} \,*}_k=
\sum _m^{  s^{\{ \lambda \} }  }T_{km}(i,\,i^\prime )
\chi ^{\{ \lambda \} \,*}_m. 
\label{tq}
\end{equation}
Note that the same matrix $T_{km}(i,\,i^\prime )$ appears in 
Eqs.(\ref{tp},\,\ref{tq}). Since the scalar product of quantum 
mechanical wave functions is invariant under the permutation 
operations, and the chosen basis is orthonormal, this matrix is 
unitary, $\sum _kT_{km}(i,\,i^\prime )T_{km^\prime }^*(i,\,i^\prime )=
\delta _{mm^\prime }$. Obviously, the functions 
$\chi ^{\{ \lambda \} }_k$ 
complex conjugate to $\chi ^{\{ \lambda \} \,*}$ also form a basis 
of an irreducible representation (a conjugate representation), and the 
transformation matrices are $T_{km}^*(i,\, i^\prime )$. 
Hence, the wave function $\Psi $ given by 
Eq.(\ref{totsym}) remains unchanged after application of operator 
$\hat{P}(i,\,i^\prime )\hat{Q}(i,\,i^\prime )$ permuting both the 
co-ordinates and spins of the pair of atoms. Also it is proved in 
the group theory that there is no way to construct the wave function 
having the Bose--Einstein symmetry property besides that given by 
Eq.(\ref{totsym}). However, the explicit form for the bosonic 
wave function in a case of an arbitrary $\{ \lambda \} $ is 
extremely lengthy. As an example, we write the formula for 
three particles, $\{ \lambda \} =\{ 2,\, 1\} $, {\it a, b} being the 
translational motion states and 1, 2 being the spin states:
\begin{eqnarray}
\Psi &=&\frac 1{3\sqrt{2}}
[\psi _a ({\bf r}_1)\psi _b({\bf r}_2)  -
\psi _a ({\bf r}_2)\psi _b({\bf r}_1)]\psi _a ({\bf r}_3) \nonumber \\
& &\times \{  \chi _1 (\sigma _1)[
\chi _1 (\sigma _2)\chi _2 (\sigma _3)  +
\chi _1 (\sigma _3)\chi _2 (\sigma _2)]  \nonumber \\ & & 
-2\chi _2 (\sigma _1) 
\chi _1(\sigma _2)\chi _1 (\sigma _3)\}  \nonumber \\ & & 
-\frac 1{3\sqrt{2}}[\psi _a ({\bf r}_2)\psi _b({\bf r}_3)  -
\psi _a ({\bf r}_3)\psi _b({\bf r}_2)]\psi _a ({\bf r}_1) \nonumber \\
& &\times \{  \chi _1 (\sigma _3)[
\chi _1 (\sigma _1)\chi _2 (\sigma _2)  +
\chi _1 (\sigma _2)\chi _2 (\sigma _1)]  \nonumber \\ & & 
-2\chi _2 (\sigma _3) 
\chi _1(\sigma _1)\chi _1 (\sigma _2)\}. 
\label{primer}
\end{eqnarray}

Consider an atomic ensemble composed of $N$ atoms in the 
state $\left| 1\right \rangle $ and $N$ atoms in the state 
$\left| 2\right \rangle $. The states $\left| 1\right \rangle $ and 
$\left| 2\right \rangle $ can be identified with the states 
$\left| F=1,\, m_F=-1\right \rangle $ and 
$\left| F=2,\, m_F=1\right \rangle $, respectively, of the ground state 
of the same bosonic isotope, e.g., $^{23}$Na or $^{87}$Rb. These states 
have equal magnetic momenta and, hence, experience the same trapping 
potential. For the sake of simplicity, we assume $N$ to be small 
enough to allow us to use a single-particle classification of 
translational motion states in the trap. Energy levels are essentially 
the levels of a single atom in the trap, the interatomic interaction 
provides only small correction to them. This assumption is not 
necessary for validity of the subsequent treatment, however, it allows 
us to present our arguments in the most compact and brief manner. 
The states of translation motion will be numbered by $\tilde j$, 
$j=0,\,1,\,2, \,...$, in the non-decreasing order in energy, so that 
$\tilde 0$ means the ground state of the trap. The tilde symbol is 
used to distinguish these states from the spin states 
$\left| 1\right \rangle $ and $\left| 2\right \rangle $. We use 
also notation $n=2N$ for the total number of atoms in the trap. 

In the standard theory of a multicomponent BEC, 
even when dealing with the 
elementary excitations of BEC, the wave function is always assumed to 
correspond to the simplest possible Young diagram characterizing both 
the co-ordinate and spin parts. This diagram has only one row consisting 
of $n$ boxes, where $n$ is the total number of atoms in the system;  
the dimension of the corresponding (totally symmetric) irreducible 
representation of the permutation group is 1, so that usually one 
writes the wave function as a product of co-ordinate and spin parts,
\begin{eqnarray} 
\Psi ({\bf r}_1,\,...\,,\,{\bf r}_n;\,\sigma _1,\,...\,,\,\sigma _n)
&=&\phi ^{\{ n \} }
({\bf r}_1,\,...\,,\,{\bf r}_n)  \nonumber  \\ & &
\times \chi ^{\{  n \}  }(\sigma _1,\,...\,,\,\sigma _n),  
\label{sovsym}
\end{eqnarray}
and the functions in the right-hand-side of this equation are invariant 
with respect to permutations of only co-ordinates or only spins. 
Now we show a way leading to creation of  complicated states corresponding 
to the general form of Eq.(\ref{totsym}). 

Initially, before the cooling process begins, atoms are thermally 
distributed. Mean occupation number for each translational state is 
much less than unity, so that we can assume that all the trap states  
occupied by the atoms are different. 

The system of thermal atoms is not in a pure state, but rather is 
described by a statistical mixture of states characterized by all the 
possible Young diagrams. The statistical weight of each Young diagram 
in this mixture is directly proportional to the number of different  
basis sets of functions those can be associated with the diagram. 
The basis sets are different in the sense that in any subspace of 
the total Hilbert space spanned by each basis there are Hilbert vectors 
not belonging to any of the remaining subspaces spanned by other 
basis sets. This is rather a subtle point. When we discussed  
earlier a question of dimensionality of an irreducible representation 
of the permutation group, we dealt with permutation of only function 
arguments within a Young diagram and assumed implicitly that each 
single-particle state $\tilde j$ was associated with a fixed box of the 
diagram. However, to determine a statistical weight of any Young 
diagram, we have to allow also permutation of state indices within 
the diagram.
As it was assumed before, the probability that a given 
state of translational motion is occupied by more than 1 atom is 
negligible before the cooling process starts. And the number of 
non-trivial ways to assign $n$ different state indices to a Young 
diagram consisting of $n$ boxes equals exactly to the number of 
non-trivial distribution of $n$ co-ordinate variables between the 
boxes of the same diagram, i.e., to $s^{\{ \lambda \} }$. 
So that for a given $\{ \lambda \} $ there are $s^{\{ \lambda \} }$ 
ways to construct different $s^{\{ \lambda \} } $-dimensional basis 
sets of co-ordinate functions. All this sets together comprise, 
according to Eq.(\ref{potom}), $n!$ functions [cf. Eq.(\ref{potom})]. 

However, not all of the Young diagrams appear in our problem. 
The restriction on possible Young diagrams 
arises from that fact that only two spin 
states are present in this Bose gas mixture. Only diagrams consisting 
of two rows, at maximum, can appear, and $\{ \lambda \} =
\{ 2N-K,K\} $, $K=0,1,2,\, ...\,,\, N$. 
If one tries to construct spin 
wave function of $N$ atoms in the state $\left| 1\right \rangle $ 
and $N$ atoms in the state $\left| 2\right \rangle $ characterized by 
a Young diagrams with 3 or more rows, the result will be identically 
zero \cite{wigner,wybourne}. Hence, only Young diagrams with not more 
than  two rows will apply also to functions of co-ordinates in our 
problem. 

There is only one non-trivial way to assign a spin state label (1 or 2) 
to each box of the diagram, so that spin effects do not increase 
the statistical weights of diagrams. 

Thus, the statistical weight of the Young diagram $\{ 2N-K,\,k\} $ 
in the initial thermal mixed state of the atomic ensemble is, 
according to Eq.(\ref{sl}) (see also Ref.\cite{wigner}), 
equals to $s^{\{ 2N-k,\, k\} }=C_{2N}^K-C_{2N}^{K-1}$, where 
$C_n^K=n!/[(n-K)!K!]$ is the binomial coefficient, and $K=
0,\,1,\, ...\,,\,N$.  The maximum of this probability distribution 
locates very close to $K=N$. Calculations show that the average 
length of the second (shorter) row of the Young diagram will be 
approximately (for $N\gg 1$) 
\begin{equation}
\left \langle K\right \rangle \approx N-\sqrt{\pi N/2}. 
\label{ksr}
\end{equation}
In other words, most probably, an observation will result in 
finding a state characterized by a Young diagram with two almost 
equal rows, which is very far from the Young diagram with only 
one row, and the probability to get the latter diagram tends to 
zero when $N$ increases. 

Then consider a cooling process. As it was said before, we assume 
a sympathetic cooling of our two-component atomic ensemble due to 
interaction with the reservoir of atoms of a different isotope kept 
at a sufficiently low temperature. 

An interaction operator acting only to atomic co-ordinates or only to 
atomic spins cannot couple a total wave function 
of $n$ identical bosons characterized by the 
Young diagram $\{ \lambda \} $ to a wave function characterized by 
$\{ \lambda ^\prime \} \neq \{ \lambda  \} $. 
Even an interatomic potential 
depending on an atomic spin orientation cannot cause such a transition 
during an elastic collision. Only if both the spin and translational 
motion states of an atom change, a transition between states 
characterized by different Young diagrams becomes possible. But such 
a process is, in fact, a spin-flip collision leading to inelastic 
losses but not facilitating the cooling process. The cooling can be 
accomplished successfully if the inelastic processes are inefficient  
on the cooling time scale. 

Thus we can see that during the sympathetic cooling process the 
type of the wave function symmetry with respect to permutations of 
co-ordinates only is conserved. This means that at the final 
stage of the cooling process an observer also will find mostly states 
characterized by Young diagrams with two rows of almost equal length. 
But the translational motion wave functions corresponding to such 
diagrams cannot be constructed using only ground state of the trap as 
a one-particle state appearing $n$ times. If one tries to construct 
a co-ordinate wave function by assigning to every box of a two-row 
diagram the same state, one will end up with identical zero. 

Therefore, the lowest energy state corresponding to the Young diagram  
$\{ 2N-K,\, K\} $ corresponds to filling of $2N-K$ boxes in the 
first row of the 
diagram with the ground one-particle state of the trap 
($\left| \tilde 0\right \rangle $) and $K$ boxes in the second row of  
the diagram with the first excited state ($\left| \tilde 1\right 
\rangle $). Thus the total energy is $2NE_{\tilde 0}+K(E_{\tilde 1}-
E_{\tilde 0})$, where $E_{\tilde j}$ is the energy of the $\tilde j$-th 
one-particle state. Usual BEC (corresponding to the Young diagram 
$\{ 2N\} $) would have the energy $2NE_{\tilde 0}$. 

Thus we see that the evaporative cooling process of a random mixture 
of the same isotope in the hyperfine states $\left| 1\right \rangle $ 
and $\left| 2\right \rangle $ with almost 100 \% probability results 
in creation of a new state of degenerate atomic Bose gas, where 
almost half of atoms are in the first excited state of the trap, and 
no further cooling can put them into the ground state. We propose 
to call such a macroscopic quantum object a Bose--Einstein--Young 
condensate (BEYC). 

We would like to stress once again that BEYC can be converted to an 
ordinary BEC only by processes where both the spin and translation 
motion of atoms is changed. But spin-flip collisions being the most 
obvious example of such a process lead to losses and heating of the 
atomic sample, due to hyperfine energy release. Spin-flip collisions, 
hence, simply destroy BEYC as well as BEC rather than facilitate 
transitions between them. 

The situation with BEYC and ordinary BEC is similar to that of 
orth- and parahydrogen molecules. Transition between these 
molecular states also requires change in the type of a Young 
diagram ($\{ 2\} $ or $\{ 1,\, 1\} $) and, hence, requires a rather 
long time even for room temperature and atmospheric presssure.

The number of atoms in the first excited trap in the BEYC state far 
exceeds both the quantum and thermal (at experimentally reachable 
temperatures) condensate depletion. The first excited state is, in 
fact, the dipole oscillation state. If the number of atoms in each of 
the two hyperfine states is well defined, the phase of these 
oscillations is indeterminate. However, it must become well-defined 
if the BEYC undergoes a continuous measurement of atomic positions, 
in close analogy with formation of an interference pattern produced by 
two counterpropagating condensates each of them was initially in the 
Fock state \cite{jy}. 

Now we have to say few words on experimental perspectives for a BEYC. 
First of all, we must answer, why BEYC has not been obtained up to now. 
In Ref.\cite{cornell1}, the evaporative cooling technique was used. 
However, $^{87}$Rb was present in that experiment in the states 
$\left| F=1,\, m_F=-1\right \rangle $ and 
$\left| F=2,\, m_F=2\right \rangle $. Their magnetic momenta differ by 
a factor 2. Hence, they experience different trapping potentials, and, 
due to the gravitation, the centers of the potential wells are displaced 
one from another. The displacement is of such a magnitude that, on the 
cooling stage, the thermal clouds of atoms in different spin states 
overlap significantly, but the two BEC do not. In fact, condensation of 
rubidium atoms takes place in different states for different 
$m_F$'s. In such a situation, it is hard to distinguish between the 
states characterized by different Young schemes, since all of them 
have the same energy, $N_{-1}E_{\tilde 0,\, -1}+N_2E_{\tilde 0,\,2}$, 
where $N_{m_F}$ is the number of atoms with the magnetic quantum 
number $m_F$ and $E_{\tilde 0,\,m_F}$ is the energy of the ground 
state of these atoms in the corresponding trapping potential influenced 
by the gravitation. 
In Ref.\cite{cornell2} an ordinary (single-component) BEC was 
prepared first. Than a fraction of the atomic ensemble was 
transferred to another hyperfine state by a two-photon 
microwave-RF pulse. Although the magnetic momenta of the two states 
were the same, creation of a BEYC was prevented by that 
circumstance that the transferred kinetic momentum (corresponding 
to the hyperfine splitting plus Zeeman shift) was negligible 
compared to the inverse trap size. Thus the translational state 
of atoms was not changed, and, hence, the wave function 
retained its symmetry properties with respect to permutation of 
only co-ordinates or only spins (i.e., remained a product of totally 
symmetric co-ordinate and spin parts). In the experiment of 
Ref.\cite{ketterle1}, the atoms were prepared initially in the 
ordinary (single-component) BEC state in a optical trap, and then 
the atomic spins were allowed to evolve due to interatomic 
interactions. The BEYC state was not reached there since it had 
a significantly higher energy than that stored in the BEC. 
The most obscure case is the case of all-optical formation of a BEC 
\cite{chapman}. In this experiment, an evaporative cooling technique 
was used, and the number of atoms in BEC was more than by order of 
magnitude smaller that the number of atoms just before the beginning 
of the cooling process. This case is beyond the scope of the present 
Rapid Communication and requires straightforward numerical (Monte Carlo) 
modeling. Only performing such a modeling, it is possible to 
understand how Young diagrams, in average, change. A scenario,  
when more than 90 \% of atoms are lost during the 
evaporative cooling process, complicated Young diagrams do not 
``survive'', and the Young diagram consisting of only one row 
of boxes prevails on the final stage, looks quite reasonable. 

Thus, to observe BEYC one needs to repeat the experiment like 
that of Ref.\cite{cornell1} (where the atoms in the two different 
hyperfine states were 
prepared {\it independently}), but with the states 
$\left| F=1,\, m_F=-1\right \rangle $ and 
$\left| F=2,\, m_F=2\right \rangle $ those have equal magnetic 
momenta. The combined magnetic trapping and gravitational potential 
will be the same, no spatial separation of condensed atoms in different 
hyperfine state will occur, and difference of BEYC from ordinary BEC 
will become apparent. An alternative approach would imply transfer 
of atoms into another hyperfine state not by a microwave-RF pulse 
as in Ref.\cite{cornell2} but by a Raman transition in two 
crossed laser beams. A non-zero angle of crossing will provide a 
kinetic momentum transfer needed for a translational motion change. 

The author thanks the NORDITA for hospitality and support 
during his visit there 
and, especially, Prof. C.J.~Pethick for stimulating discussions. The 
author thanks also Prof. D.A.~Varshalovich for helpful comments. 
The financial support from the NWO, grant NWO--047--009.010, 
the RFBR, grant 02--02--17686, 
and the Ministry of Education of Russia, grant UR.01.01.040, is  
acnowledged.

\end{document}